\documentclass[aps,prl,twocolumn,amsmath,amssymb,showpacs,superscriptaddress,notitlepage,longbibliography]{revtex4-1}
\usepackage[colorlinks=true,linkcolor=blue,anchorcolor=red,citecolor=blue, urlcolor=blue]{hyperref}
\usepackage{bm}
\usepackage{graphicx}
\usepackage{color}

\begin{document}

\title{Band signatures for strong nonlinear Hall effect in bilayer WTe$_2$}

\author{Z. Z. Du}
\affiliation{Shenzhen Institute for Quantum Science and Engineering and Department of Physics, Southern University of Science and Technology, Shenzhen 518055, China}
\affiliation{Shenzhen Key Laboratory of Quantum Science and Engineering, Shenzhen 518055, China}
\affiliation{School of Physics, Southeast University, Nanjing 211189, China}

\author{C. M. Wang}
\affiliation{Department of Physics, Shanghai Normal University, Shanghai 200234, China}
\affiliation{Shenzhen Institute for Quantum Science and Engineering and Department of Physics, Southern University of Science and Technology, Shenzhen 518055, China}
\affiliation{Shenzhen Key Laboratory of Quantum Science and Engineering, Shenzhen 518055, China}

\author{Hai-Zhou Lu}
\email{Corresponding author: luhz@sustc.edu.cn}
\affiliation{Shenzhen Institute for Quantum Science and Engineering and Department of Physics, Southern University of Science and Technology, Shenzhen 518055, China}
\affiliation{Shenzhen Key Laboratory of Quantum Science and Engineering, Shenzhen 518055, China}

\author{X. C. Xie}
\affiliation{International Center for Quantum Materials, School of Physics, Peking University, Beijing 100871, China}
\affiliation{Collaborative Innovation Center of Quantum Matter, Beijing 100871, China}

\date{\today }

\begin{abstract}
Unconventional responses upon breaking discrete or crystal symmetries open avenues for exploring emergent physical systems and materials. By breaking inversion symmetry, a nonlinear Hall signal can be observed, even in the presence of time-reversal symmetry, quite different from the conventional Hall effects. Low-symmetry two-dimensional materials are promising candidates for the nonlinear Hall effect, but it is less known when a strong nonlinear Hall signal can be measured, in particular, its connections with the band-structure properties.
By using model analysis, we find prominent nonlinear Hall signals near tilted band anticrossings and band inversions. These band signatures can be used to explain the strong nonlinear Hall effect in the recent experiments on two-dimensional WTe$_{2}$. This Letter will be instructive not only for analyzing the transport signatures of the nonlinear Hall effect but also for exploring unconventional responses in emergent materials.
\end{abstract}

\maketitle

\begin{figure}[htpb]
\centering
  \includegraphics[width=0.36\textwidth]{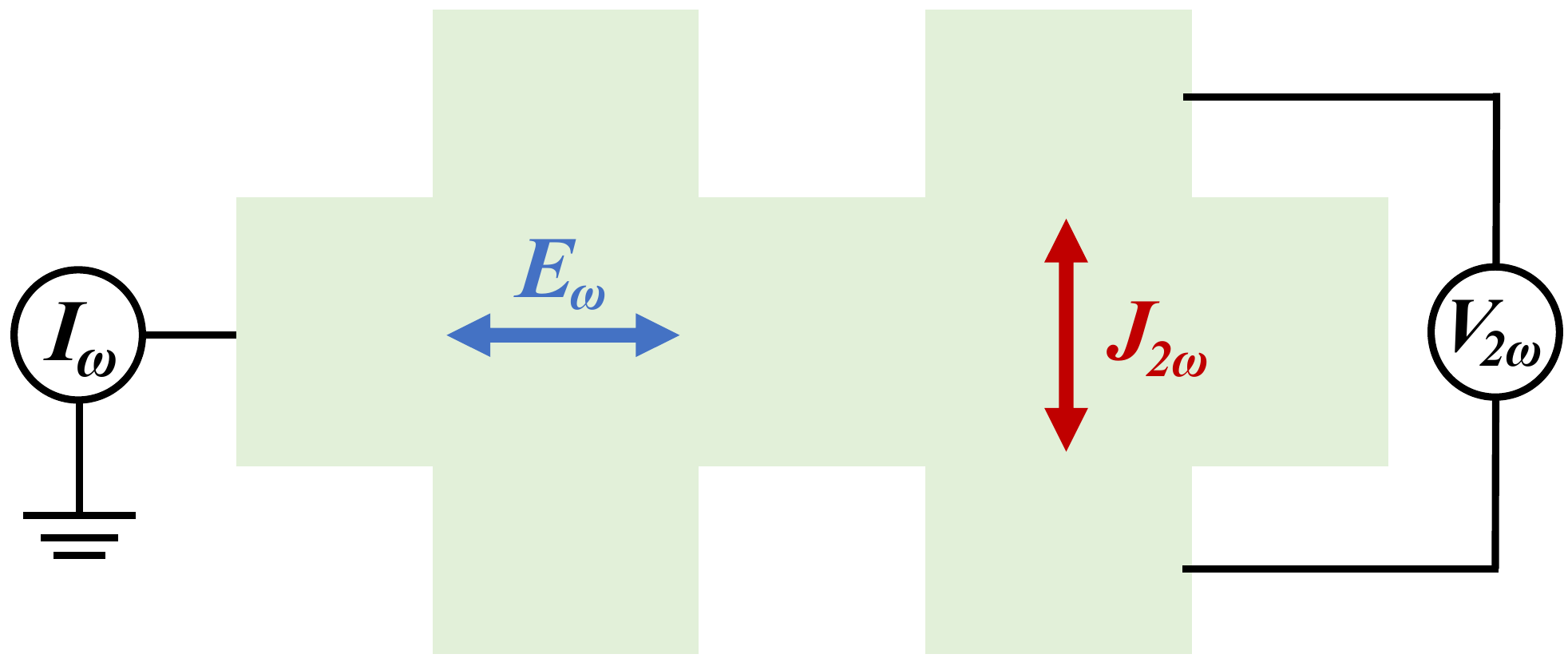}\\
  \caption{Schematic of how to measure the nonlinear Hall effect in a standard Hall bar. The experimentally measured $I-V$ relation is related to the theoretical Berry dipole defined by the electric field-current density ($E-J$) relation. }\label{Fig:NLHE}
\end{figure}

{\color{blue}\emph{Introduction.}}--
The Hall effects are among the most paradigmatic phenomena in condensed matter physics because of their deep connections with the geometry and topology \cite{Klitzing80prl,QHE2012,Xiao10rmp}.
All known measurable Hall effects need magnetic fields or magnetic dopants to break time-reversal symmetry \cite{Klitzing80prl,QHE2012,Nagaosa10rmp,yasuda2016NP}.
Recently, a new Hall effect, i.e., the nonlinear Hall effect, was proposed \cite{Sodemann15prl}, which does not need time-reversal symmetry breaking, but inversion symmetry breaking. To measure it, the zero- or double-frequency component of the Hall conductance is rectified in response to a driving electric field oscillating at a low frequency (Fig.~\ref{Fig:NLHE}). This measurement is highly accessible to known experimental conditions, and thus opens new avenues of probing the spectral, symmetry, and topological properties of a number of emergent materials. More importantly, the idea can be generalized to other unconventional responses upon breaking discrete and crystal symmetries, leading to a promising and unknown territory.
Despite the symmetry argument, how a considerably strong nonlinear Hall signal can be measured remains less known and is the focus of recent explorations.

\begin{figure}[htpb]
\centering
  \includegraphics[width=0.45\textwidth]{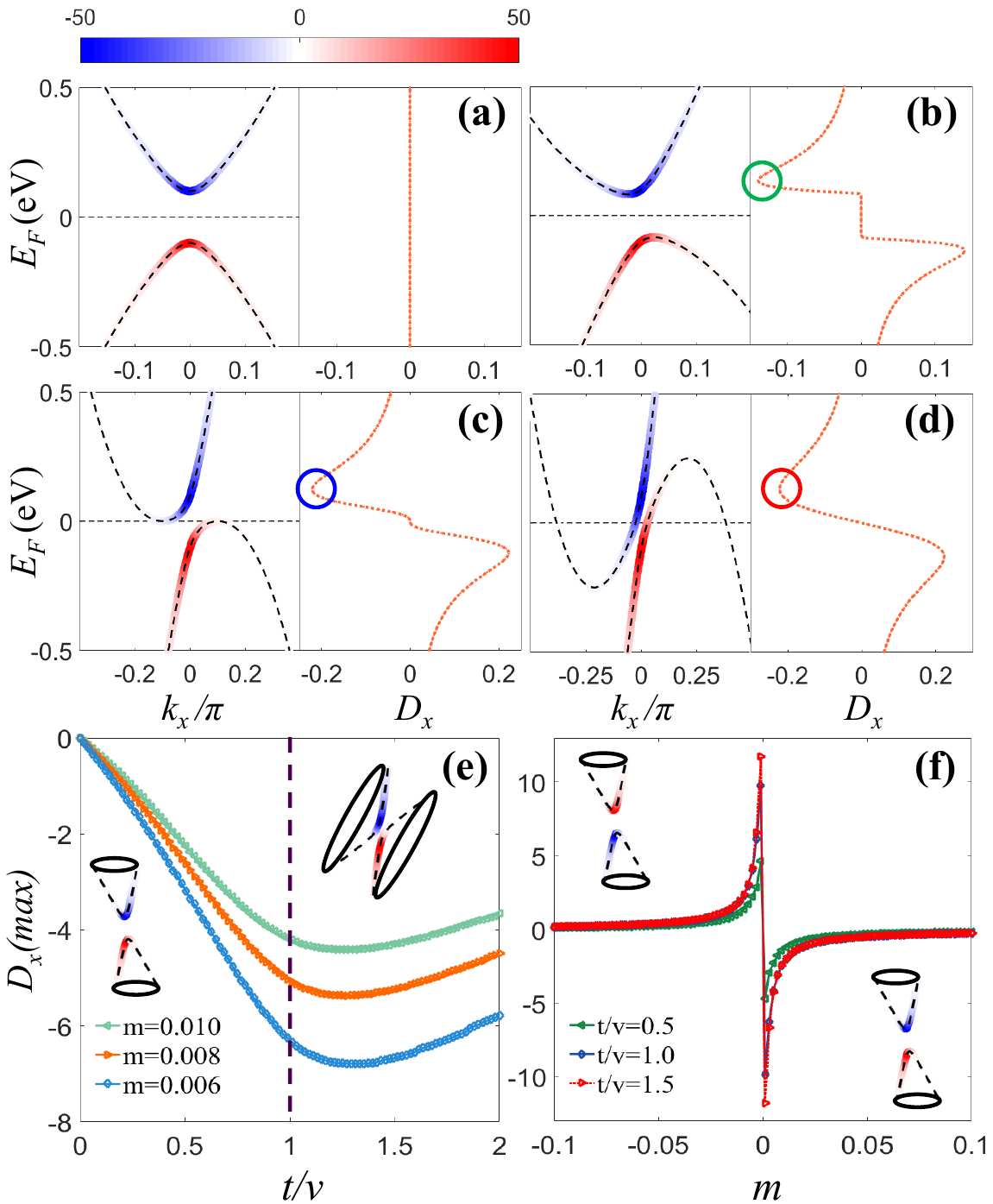}\\
  \caption{(a)-(d) (left panels) The band structure of the tilted Dirac fermion model  in Eq. \eqref{Eq:TiltedDirac} as a function of $k_x$ at $k_y=0$ with the tilt $t/v=0$ (a), $t/v=0.5$ (b), $t/v=1.0$ (c) and $t/v=1.5$ (d). The color bar is for the Berry curvature, whose intensity is plotted on each band. (a)-(d) (right panels) The corresponding Berry dipole $D_x\equiv D_{xz}$, measured by the driving current along the $x$ direction and the Hall voltage along the $y$ direction.
  (e), (f) The maximum of the Berry dipole $D_x$ [the circles in (b)-(d)] as a function of the tilt $t/v$ for different mass gap (e) and as a function of the gap $m$ for different tilt angle (f). The other parameters are $v=1~\mathrm{eV {\AA}}$, $\alpha=1~\mathrm{eV {\AA}^{2}}$, $\eta=-1$, and $m=0.2~\mathrm{eV}$. }\label{Fig:DiracCone}
\end{figure}

\begin{figure}[htpb]
\centering
  \includegraphics[width=0.42\textwidth]{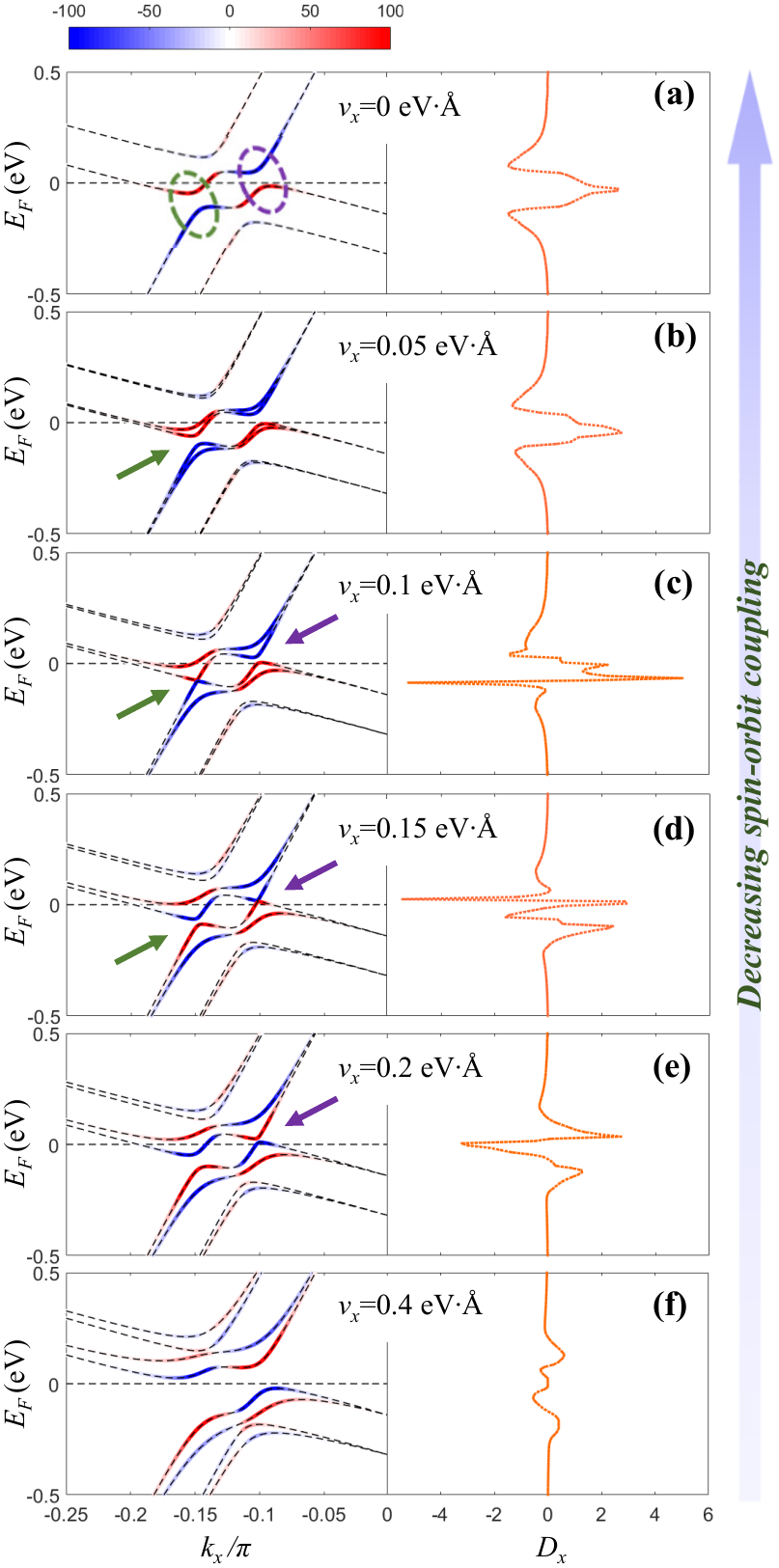}\\
  \caption{(Left panels) The band structure of the coupled Dirac Fermion model as a function of $k_x$ at $k_y=0$ in Eq.~\eqref{Eq:Bilayer} with increasing spin-orbit coupling strength ($\nu_x$). The ovals mark the titled Dirac cones. The arrows indicate the band inversion and anticrossings. (Right panels) The Berry dipole $D_x\equiv D_{xz}$ for corresponding spin-orbit coupling strength. Here $\nu_{y}=0~\mathrm{eV {\AA}}$, $\gamma=0.05~\mathrm{eV}$ and the other parameters in $\hat{\mathcal{H}}_{d_{i}}$ are $K_{1}=0.1\pi$, $K_{2}=0.15\pi$, $v_{1}=v_{2}=2~\mathrm{eV {\AA}}$, $t_{1}=t_{2}=1.5~\mathrm{eV {\AA}}$, $m_{1}=m_{2}=0.1~\mathrm{eV}$, $\eta_{1}=-1$, $\eta_{2}=1$, $E_{1}=0.02~\mathrm{eV}$, $E_{2}=-0.08~\mathrm{eV}$.}\label{Fig:WTe}
\end{figure}

\begin{figure}[htpb]
\centering
  \includegraphics[width=0.45\textwidth]{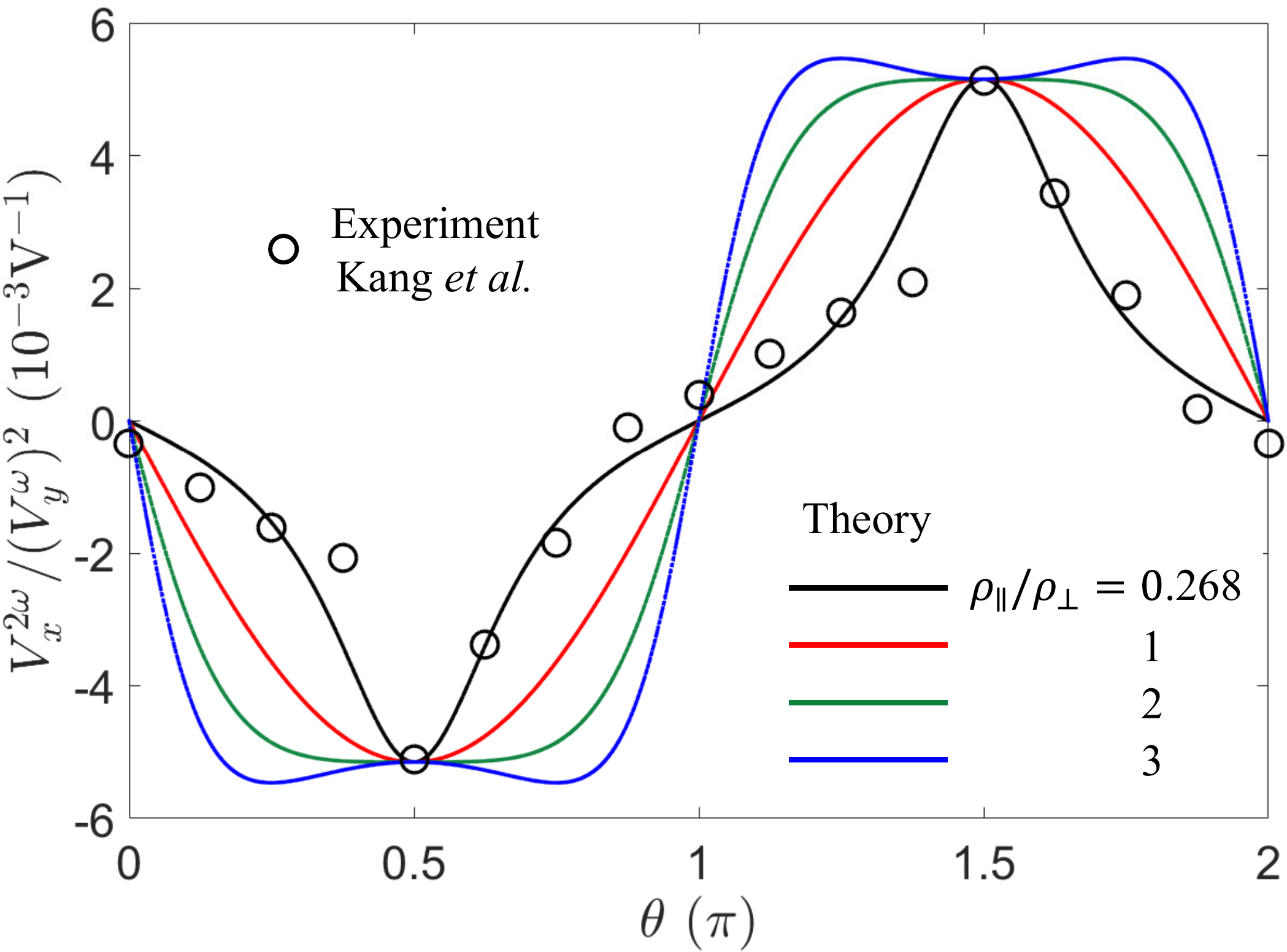}\\
  \caption{The angular dependence of the nonlinear Hall voltage as a function of the angle $\theta$ between the driving current and principal axes for different resistance anisotropy $\rho_{||}/\rho_{\perp}$. The circles are the experimental data adapted from \cite{Kang18arXiv}. $\rho_{||}/\rho_{\perp}=0.268$ is read out from the experiment. For contrast, different curves are scaled so that they cross with the experimental data at $\theta=\pi/2,~3\pi/2$, where
  $V^{2\omega}_{x}/(V^{\omega}_{y})^{2}\simeq5.15\times10^{-3}V^{-1}$.}\label{Fig:Angle}
\end{figure}

In this Letter, we investigate the relation between the nonlinear Hall effect and signatures of energy bands in 2D systems. The nonlinear Hall response is proportional to the so-called Berry dipole \cite{Sodemann15prl}. By analyzing a generic model of tilted massive Dirac fermions, we find that the Berry dipole is strong near tilted band anticrossings and band inversions (Fig.~\ref{Fig:DiracCone}). To give an example of broad interest, we calculate the nonlinear Hall response in the bilayer WTe$_{2}$ (Fig.~\ref{Fig:WTe}), in which spin-orbit coupling modulates band anticrossings and band inversions to give divergences of the Berry dipole and strong nonlinear Hall responses. Our results are consistent with the strong nonlinear Hall response and its angular dependence (Fig.~\ref{Fig:Angle}) in recent experiments \cite{Ma18nat,Kang18arXiv} and will inspire more experiments on the nonlinear Hall effects and related novel phenomena.

{\color{blue}\emph{Review of the nonlinear Hall effect.}} --
The nonlinear Hall effect originates from the
dipole moment of the Berry curvature in momentum space, or, in short, the Berry dipole.
When applying an oscillating electric field $\mathbf{E}(t)=\mathrm{Re}\{\mathcal{E}e^{i\omega t}\}$ with the amplitude vector $\mathcal{E}$ and frequency $\omega$, the nonlinear current in response to the electric field can be formally decomposed into the dc and double-frequency components $J_{a}=\mathrm{Re}\big\{J^{(0)}_{a}+J^{(2)}_{a}e^{i2\omega t}\big\}$, with $J^{(0)}_{a}=\chi^{(0)}_{abc}\mathcal{E}_{b}\mathcal{E}^{*}_{c}$ and $J^{(2)}_{a}=\chi^{(2)}_{abc}\mathcal{E}_{b}\mathcal{E}_{c}$, respectively.
For time-reversal symmetric systems, their nonlinear Hall coefficients have the form \cite{Sodemann15prl}
\begin{eqnarray}
\chi^{(0)}_{abc}=\chi^{(2)}_{abc}=\varepsilon^{acd}D_{bd}e^{3}\tau/[2\hbar^{2}(1+i\omega\tau)],
\end{eqnarray}
where $-e$ is the electron charge and $\tau$ is the momentum relaxation time. According to the experiments \cite{Ma18nat,Kang18arXiv}, the typical frequency of the ac electric field is about 10-1000 Hz and $\tau$ is approximately picoseconds; thus the frequency dependence in the denominator can be neglected due to $\omega\tau\ll 1$, which is the key difference from nonlinear optics.
The Berry dipole can be found as
\begin{eqnarray}\label{Dipole}
D_{bd}
&=&-\sum_{i}\int\frac{d^{d}k}{(2\pi)^{d}}\frac{\partial \epsilon^i_{\mathbf{k}}}{\partial k_b}\Omega^{d}_{i\textbf{k}}
\frac{\partial f_{\textbf{k}}}{\partial\epsilon^i_{\mathbf{k}}},
\end{eqnarray}
where $\varepsilon^{abc}$ is the Levi-Civita symbol, $a,b,c,d\in\{x,y,z\}$, $f_{\textbf{k}}$ is the Fermi distribution, and the Berry curvature can be found as \cite{Xiao10rmp}
\begin{eqnarray}
\Omega^{a}_{i \textbf{k}}
&=&-2\varepsilon^{abc}\sum_{j\neq i}\frac{\mathrm{Im}\langle i|\partial\hat{\mathcal{H}}/\partial k_{b}|j\rangle\langle j|\partial\hat{\mathcal{H}}/\partial k_{c}|i\rangle}{(\epsilon^{i}_{\mathbf{k}}-\epsilon^{j}_{\mathbf{k}})^{2}},
\end{eqnarray}
where $|i\rangle$ refers to the eigenstate in band $i$ with energy $\epsilon^{i}_{\mathbf{k}}$ for a given wave vector $\mathbf{k}$.
The derivative of the Fermi distribution in this definition of the Berry dipole infers that the states close to the Fermi surface mainly contribute to the nonlinear Hall response.
Note that, under time reversal,  $\Omega^{a}_{i}(-\textbf{k})=-\Omega^{a}_{i}(\textbf{k})$ and $\partial\epsilon^i_{\mathbf{k}}/\partial (-k_b)=-\partial\epsilon^i_{\mathbf{k}}/\partial k_b$; thus the integral in Eq.~\eqref{Dipole} can survive in time-reversal symmetric systems. Meanwhile, inversion symmetry must be broken to support finite Berry curvature so that the integral in Eq.~\eqref{Dipole} does not vanish.
Table \ref{Tab:Comparison} compares the differences between the linear and nonlinear Hall effects.

\begin{table}[htb]
\centering
\caption{Comparison of the linear and nonlinear Hall effects. The Berry dipole in Eq.~(\ref{Dipole}) can be written as a Fermi-sea integral by using the integration by parts.}\label{Tab:Comparison}
\begin{ruledtabular}
\begin{tabular}{ccc}
& Linear  & Nonlinear \\
Broken symmetry & Time reversal  & Inversion \\ I-V relation & $\sim V^y_{\omega}/I^x_{\omega}$ & $\sim V^y_{2\omega}/I^x_{\omega}I^x_{\omega}$ \\
Fermi-sea integral of & Berry curvature $ \Omega_{\mathbf{k}} $  & $\partial \Omega_{\mathbf{k}} / \partial \mathbf{k}$ \\
In-gap signal & Possible (e.g., QHE)  & No  \\
Angular dependence & No & Eq.~(\ref{Eq:PHE}) \\
\end{tabular}
\end{ruledtabular}
\end{table}

{\color{blue}\emph{Tilted 2D massive Dirac model.}}-- According to Eq.~(\ref{Dipole}), the nonlinear Hall effect requires anisotropic bands with finite Berry curvature, so that $\Omega(\mathbf{k})\neq \Omega(-\mathbf{k})$. The tilted 2D massive Dirac model can give the ingredients
\begin{eqnarray}\label{Eq:TiltedDirac}
\hat{\mathcal{H}}_{d}=tk^{x}+v(k^{y}\sigma_{x}+\eta k^{x}\sigma_{y})+(m/2-\alpha k^{2})\sigma_{z},
\end{eqnarray}
where $(k_{x}, k_{y})$ are the wave vectors, $(\sigma_{x}, \sigma_{y}, \sigma_{z})$ are the Pauli matrices, $\eta=\pm1$, $m$ is the gap, and $t$ tilts the Dirac cone along the $x$ direction.  Compared to Ref. \cite{Sodemann15prl}, $\alpha$ is introduced to regulate topological properties as $k\rightarrow \infty$ \cite{Shen17book}. The time reversal of the model contributes equally to the Berry dipole, so it is enough to study this model only. The model describes two energy bands (denoted as $\pm$) with the band dispersions $\varepsilon^{\pm}_{\mathbf{k}}=tk_{x}\pm[v^{2}k^{2}+(m/2-\alpha k^{2})^{2}]^{1/2}$, where $k^{2}\equiv k^{2}_{x}+k^{2}_{y}$.
In the $x-y$ plane, the Berry curvature behaves like a pseudoscalar, with only the $z$ component $\Omega^z_{\pm \mathbf{k}}=\pm\eta v^{2}(m/2-\alpha k^{2})/\{2[v^{2}k^{2}+(m/2-\alpha k^{2})^{2}]^{3/2}\}$.
Correspondingly, the Berry dipole behaves as a pseudovector constrained in the $x-y$ plane, e.g., $D_{xz}\rightarrow D_{x}$ and $D_{yz}\rightarrow D_{y}$.

{\color{blue}\emph{Berry dipole near tilted band anticrossing}.} -- The band dispersion, Berry curvature, and Berry dipole of the Dirac cone in Eq.~(\ref{Eq:TiltedDirac}) is shown in Fig.~\ref{Fig:DiracCone}. If there is no tilt ($t=0$), Fig.~\ref{Fig:DiracCone}(a) shows that the Berry curvature is symmetrically concentrated around the band edges so there is no Berry dipole.
As the Dirac cone is tilted, the Berry curvature no longer symmetrically distributes, so the peaks of the Berry dipole appear near the band edges, as shown in Fig.~\ref{Fig:DiracCone}(b)-(d).
Here the tilt along the $x$ direction does not break the $y$-direction mirror reflection symmetry ($k_{y}\rightarrow-k_{y}$), so only the $x$ component of the Berry dipole is nonzero. Figure~\ref{Fig:DiracCone}(e) shows that the Berry dipole vanishes at zero tilt ($t=0$), grows with increasing tilt, and reaches a maximum near the critical point ($t=v$) beyond which the Dirac cone is overtilted ($t>v$).

The Berry dipole vanishes in the gap due to the absence of carriers. Thus, unlike the Berry curvature in the intrinsic anomalous Hall effect, the Berry dipole does not build up across the band gap in the nonlinear Hall effect. The Berry dipole of the conduction and valence bands have opposite signs, because the bands of the model are particle-hole symmetric and the Berry curvature changes sign across the gap. The Berry curvature usually reaches maximum at the band edges, but the Berry dipole peaks do not lie exactly at the band edges because the group velocity vanishes at the band edges (see details in Sec. SII of \cite{Supp}).

{\color{blue}\emph{Divergence of Berry dipole near band inversions.}}--
By varying $m$ in model \eqref{Eq:TiltedDirac} from the same sign as $\alpha$ to the opposite sign, the system undergoes a band inversion, accompanied by a topological phase transition from the quantum anomalous Hall insulator to a trivial 2D insulator \cite{Bernevig06sci,Lu10prb}.
During the transition, the Berry curvature of the conduction and valence bands exchanges sign \cite{Lu13prl-QAH} and the linear Hall conductance changes by $e^2/h$.
Consequently, one can expect that the Berry dipole also changes sign during this topological phase transition.
Figure~\ref{Fig:DiracCone}(f) shows that divergences appear at the critical point where $m$ changes sign.
The divergences near the gap closing indicate that the nonlinear Hall response can be significantly enhanced near the critical point of the topological phase transition. The divergences of the Berry dipole can serve as a probe for the topological phase transition of 2D tilted Dirac fermions.

{\color{blue}\emph{Bilayer WTe$_{2}$.}}-- Now we show that the above-mentioned  band signatures, i.e., the tilted band anticrossing and band inversion, can give strong nonlinear Hall signals in bilayers of WTe$_{2}$.
According to the above analysis, a time-reversal symmetric system needs inversion symmetry breaking to have large Berry curvature and large band tilt near the Fermi surface, in order to show a strong nonlinear Hall response.
A number of materials meet the requirements \cite{wang2016PRL,autes2016PRL,ma2017NP,zhang2018TDM,Facio18PRL}.
Among them, the most promising candidate is WTe$_{2}$, which has attracted tremendous attention due to its properties associated with a significant distribution of the Berry curvature  \cite{ali2014Nature,soluyanov2015Nature,wang2016NC,qian2014Science,zheng2016AM,fei2017NP,tang2017NP,wu2018Science,xu2018NP,macneill2017NP,macneill2017PRB,fei2018Nature}, e.g., the bulk WTe$_{2}$ is a type-II Weyl semimetal \cite{soluyanov2015Nature,wang2016NC},
its monolayer can host the high-temperature quantum spin Hall effect \cite{qian2014Science,zheng2016AM,fei2017NP,tang2017NP,wu2018Science} and electrically switchable circular photogalvanic effect \cite{xu2018NP}. For the nonlinear Hall effect, the WTe$_{2}$ bilayer has several advantages. It can be naturally exfoliated from bulk crystals. It has no inversion symmetry \cite{macneill2017NP,macneill2017PRB}, while monolayers require gating to break inversion symmetry. It exhibits a spontaneous out-of-plane electric polarization \cite{fei2018Nature}, which can help tuning spin-orbit coupling through gating.

In the absence of spin-orbit coupling, the bilayer WTe$_{2}$ can be described by four tilted Dirac fermions \cite{Muechler2016}.
In the presence of spin-orbit coupling, we propose an effective model of the bilayer WTe$_{2}$, which couples four pairs of tilted Dirac cones and reads
\begin{eqnarray}\label{Eq:Bilayer}
\hat{\mathcal{H}}_{\mathrm{B}}=\left(
                        \begin{array}{cccc}
                          \hat{\mathcal{H}}_{d_{1}} & \hat{\mathcal{P}} & 0 & \gamma \\
                          \hat{\mathcal{P}} & \hat{\mathcal{H}}_{d_{1}} & \gamma & 0 \\
                          0 & \gamma & \hat{\mathcal{H}}_{d_{2}} & \hat{\mathcal{P}} \\
                          \gamma & 0 & \hat{\mathcal{P}} & \hat{\mathcal{H}}_{d_{2}} \\
                        \end{array}
                      \right),
\end{eqnarray}
where $\hat{\mathcal{H}}_{d_{i}}$ describes the tilted Dirac cone located at $\mathbf{K}_{i}$ (see Sec. SIII of \cite{Supp}),
\begin{eqnarray}
\hat{\mathcal{P}}=\left(
                        \begin{array}{cc}
                          \nu_{x}k_{x}-i\nu_{y}k_{y} & 0 \\
                          0 & -\nu_{x}k_{x}-i\nu_{y}k_{y} \\
                        \end{array}
                      \right),
\end{eqnarray}
and $\nu_{x}$ and $\nu_{y}$ measure the spin-orbit coupling strength along the $x$ and $y$ directions, respectively, and can be tuned by a gate voltage applied along the $z$ direction.
This model describes only the negative half of the Brillouin zone, which contributes equally to the Berry dipole as the positive half does.
This model can qualitatively capture many key features of the bilayer WTe$_{2}$ including the band structure evolution with spin-orbit coupling and the nonlinear Hall effect.

The evolution of the band structure with spin-orbit coupling is shown in the left panels of Fig.~\ref{Fig:WTe}.
According to the first-principles calculations \cite{Muechler2016}, in the absence of spin-orbit coupling the bilayer WTe$_{2}$ is a semimetal with a gap opened by the interlayer coupling, and each band is twofold degenerate. We can identify two tilted Dirac cones (marked by the ovals) in Fig.~\ref{Fig:WTe}. The Berry curvature of the two Dirac cones are opposite because the two layers of the bilayer WTe$_{2}$ are related through a mirror reflection. The experiments have observed that the bilayer WTe$_{2}$ becomes insulating at low temperatures, implying a possible gap formation due to spin-orbit coupling \cite{fei2017NP}. As spin-orbit coupling is turned on, the degeneracy of the bands is lifted. The band structure shown in Fig.~\ref{Fig:WTe}(f) is well gapped due to spin-orbit coupling and is likely the case in the experiments. As spin-orbit coupling decreases, Fig.~\ref{Fig:WTe} shows that the band gaps shrink and the system undergos two band inversions [indicated by the arrows in Fig.~\ref{Fig:WTe}(b) $\leftrightarrow$ (d) and (c) $\leftrightarrow$ (e)], at which the conduction band exchanges the sign of the Berry curvature with the valence band.
This band structure evolution is consistent with the first-principles calculations and experiments \cite{Muechler2016,fan2015CPB}.

The corresponding Berry dipole results of these band structures are shown in the right panels of Fig.~\ref{Fig:WTe}. As spin-orbit coupling decreases, the magnitude of the nonlinear Hall response increases with the shrinking gaps. In particular, the divergences appear at two band inversions. These divergences at the band inversions follow the mechanism shown in Fig.~\ref{Fig:DiracCone}(f). The Berry dipole behavior is qualitatively consistent with the giant nonlinear Hall effect observed in the experiments, in which the nonlinear Hall response shows many sign changes and sharp resonancelike peaks as the gate voltage aligns the Fermi energy with the band signatures, such as the tilted band anticrossings and band inversions \cite{Ma18nat}.

{\color{blue}\emph{Angular dependence of the nonlinear Hall voltage.}}--
For systems without highfold rotation axis perpendicular to the 2D plane, it is known that the linear longitudinal and transverse resistances show twofold angular dependence as $\rho_{xx}=\rho_{\parallel}\cos^{2}\theta+\rho_{\perp}\sin^{2}\theta$ and $\rho_{xy}=(\rho_{\parallel}-\rho_{\perp})\sin\theta\cos\theta$.
Here $\rho_{\parallel}$, $\rho_{\perp}$ refer to the resistivity along the principal axes and $\theta$ is the angle between the driving current and the $||$ principal axis.
The nonlinear Hall coefficients also show angular dependence, but different from the linear case, the angular dependence is onefold
\begin{eqnarray}\label{Eq:PHE}
\chi_{xxy}&=&-\chi_{yxx}=\chi_{\parallel}\cos\theta-\chi_{\perp}\sin\theta,
\nonumber\\
\chi_{xyy}&=&-\chi_{yyx}=\chi_{\parallel}\sin\theta+\chi_{\perp}\cos\theta,
\end{eqnarray}
where $\chi_{\parallel}$ and $\chi_{\perp}$ represent the nonlinear Hall coefficients along the principal axes (see details in Sec. SIV of \cite{Supp}).
All these coefficients change sign when the current direction is reversed.
The experiments measure only the Hall voltages, so we transform the angular dependence from the nonlinear Hall coefficients to that of the Hall voltages, and for 2D WTe$_{2}$ it reads (see Sec. SIV of \cite{Supp})
\begin{eqnarray}
\frac{V^{2\omega}_{x}}{(V^{\omega}_{y})^{2}}
&=&\frac{\chi_{\parallel}\rho^{2}_{\parallel}\rho_{\perp}\sin\theta}
{(\rho_{\parallel}\sin^{2}\theta+\rho_{\perp}\cos^{2}\theta)^{2}},
\end{eqnarray}
where we have assumed that the driving current is along the $y$ direction and the nonlinear Hall voltage is measured along the $x$ direction.

This means that the line shape of the angular dependence of $ V^{2\omega}_{x}/(V^{\omega}_{y})^{2}$ depends on the resistance anisotropy ratio $\rho_{\parallel}/\rho_{\perp}$, as shown in Fig.~\ref{Fig:Angle}.
For $\rho_{\parallel}/\rho_{\perp}=0.268$, we can quantitatively reproduce the case measured in the experiment \cite{Kang18arXiv}.
In the isotropic case ($\rho_{\parallel}/\rho_{\perp}=1$), the angular dependence is simply a sine function.
For $\rho_{\parallel}/\rho_{\perp}>1$, the peaks of the sine function are flattened, which is not the case in the experiment \cite{Kang18arXiv}.
If the resistance anisotropy ratio is large enough, say $\rho_{\parallel}/\rho_{\perp}=3$, the maxima at $\theta=\pi/2,~3\pi/2$ turn into minima, giving a double-peak line shape. This double-peak angular dependence can be observed by modifying the resistance anisotropy through tuning the gate voltages \cite{Ma18nat}.

We thank insightful discussions with Huimei Liu.
This work was supported by the Guangdong Innovative and Entrepreneurial Research Team Program (2016ZT06D348), the National Basic Research Program of China (2015CB921102), the National Key R \& D Program (2016YFA0301700), the
National Natural Science Foundation of China (11534001, 11474005, 11574127), and the Science, Technology and Innovation Commission of Shenzhen Municipality (ZDSYS20170303165926217, JCYJ20170412152620376).


%

\end{document}